\documentclass[sn-mathphys,Numbered,iicol]{sn-jnl}

\usepackage{graphicx}%
\usepackage{multirow}%
\usepackage{amsmath,amssymb,amsfonts}%
\usepackage{amsthm}%
\usepackage{mathrsfs}%
\usepackage[title]{appendix}%
\usepackage{xcolor}%
\usepackage{textcomp}%
\usepackage{manyfoot}%
\usepackage{booktabs}%
\usepackage{algorithm}%
\usepackage{algorithmicx}%
\usepackage{algpseudocode}%
\usepackage{listings}%
\usepackage{bbm}

\theoremstyle{thmstyleone}%
\theoremstyle{thmstyletwo}%

\theoremstyle{thmstylethree}%

\newcounter{multieqs}

\newcommand{\be}{\begin{equation}}
	\newcommand{\ee}{\end{equation}}
\newcommand{\eq}[1]{(\ref{#1})}

\def\bd{\begin{document}}
	\def\ed{\end{document}}
\def\nn{\nonumber}
\def\bea{\begin{eqnarray}}
	\def\eea{\end{eqnarray}}

\def\la{\langle}
\def\ra{\rangle}

\def\N{{\cal N}}
\def\sst{\scriptscriptstyle}
\def\thetabar{\bar\theta}
\def\Tr{{\rm Tr}}
\def\one{\mbox{1 \kern-.59em {\rm l}}}

\def\a{\alpha}      \def\da{{\dot\alpha}}  \def\dA{{\dot A}}
\def\b{\beta}       \def\db{{\dot\beta}}  
\def\g{\gamma}  \def\G{\Gamma}  \def\dc{{\dot\gamma}}  
\def\d{\delta}  \def\D{\Delta}  \def\ddt{\dot\delta}  
\def\e{\epsilon}        \def\ve{\varepsilon}  
\def\f{\phi}    \def\F{\Phi}    \def\vvf{\f}  
\def\h{\eta}  
\def\k{\kappa}  
\def\l{{\lambda}} \def\L{\Lambda}  
\def\m{\mu} \def\n{\nu}  
\def\o{\omega}  
\def\p{\pi} \def\P{\Pi}  
\def\r{\rho}  
\def\s{\sigma}  \def\S{\Sigma}  
\def\t{\tau}  
\def\th{\theta} \def\Th{\Theta} \def\vth{\vartheta}  
\def\X{\Xeta}  
\def\z{\zeta}  

\def\na{\nabla}  

\def\cA{{\cal A}} \def\cB{{\cal B}} \def\cC{{\cal C}}  
\def\cD{{\cal D}} \def\cE{{\cal E}} \def\cF{{\cal F}}  
\def\cG{{\cal G}} \def\cH{{\cal H}} \def\cI{{\cal I}}  
\def\cJ{{\cal J}} \def\cK{{\cal K}} \def\cL{{\cal L}}  
\def\cM{{\cal M}} \def\cN{{\cal N}} \def\cO{{\cal O}}  
\def\cP{{\cal P}} \def\cQ{{\cal Q}} \def\cR{{\cal R}}  
\def\cS{{\cal S}} \def\cT{{\cal T}} \def\cU{{\cal U}}  
\def\cV{{\cal V}} \def\cW{{\cal W}} \def\cX{{\cal X}}  
\def\cY{{\cal Y}} \def\cZ{{\cal Z}}

\def\Ah{{\hat{A}}}  
\def\Dh{{\hat{D}}}
\def\Gh{{\hat{G}}}
\def\Fh{{\hat{F}}}
\def\Ih{{\hat{I}}} 
\def\Jh{{\hat{J}}} 
\def\Kh{{\hat{K}}}
\def\Lh{{\hat{L}}} 
\def\Ph{{\hat{P}}}
\def\Rh{{\hat{R}}}
\def\Vh{{\hat{V}}} 
\def\Xh{{\hat{X}}}

\def\ah{{\hat{a}}}
\def\bh{{\hat{b}}}
\def\ch{{\hat{c}}}
\def\gh{{\hat{g}}}
\def\dh{{\hat{d}}}
\def\hh{{\hat{h}}}
\def\uh{{\hat{u}}}  
\def\vh{{\hat{v}}}
\def\xh{{\hat{x}}} 
\def\yh{{\hat{y}}}
\def\zh{{\hat{z}}}
\def\ph{{\hat{p}}}
\def\qh{{\hat{q}}}
\def\thh{{\hat{t}}}  
\def\xih{\hat{\xi}}  
\def\Psih{\hat{\Psi}}    
\def\mh{{\hat{m}}}
\def\nh{{\hat{n}}}
\def\ih{{\hat{i}}}
\def\jh{{\hat{j}}}
\def\kh{{\hat{k}}}
\def\aah{{\hat{\alpha}}}
\def\bbh{{\hat{\beta}}}
\def\ggh{{\hat{\gamma}}}
\def\llh{{\hat{\ell}}} 
\def\ph{{\hat{p}}}

\def\psit{\tilde{\psi}}  
\def\Psit{\tilde{\Psi}}   
\def\Psibt{\tilde{\bar{Psi}}}  

\def\st{\tilde{\sigma}}  

\def\delt{\tilde{\delta}}
\def\Phit{\tilde{\Phi}}   
\def\Phitb{\overline{\tilde{Phi}}}  
\def\tht{\tilde{\th}}  
\def\lt{\tilde{\l}}
\def\chit{\tilde{\chi}}   
\def\phit{\tilde{\phi}} 

\def\Ft{\tilde{F}}
\def\Kt{\tilde{K}}

\def\At{\tilde{A}}
\def\Bt{\tilde{B}}
\def\Ct{\tilde{C}}
\def\Dt{\tilde{D}}
\def\Et{\tilde{E}}
\def\Ht{\tilde{H}}
\def\It{\tilde{I}}
\def\Jt{\tilde{J}}
\def\Qt{\tilde{Q}}  
\def\Rt{\tilde{R}}  
\def\Mt{\tilde{M }}  
\def\Nt{\tilde{N}}   
\def\St{\tilde{S}}
\def\Vt{\tilde{V}}
\def\Xt{\tilde{X}} 
\def\at{\tilde{a}}
\def\ct{\tilde{c}}
\def\dt{\tilde{d}}
\def\htt{\tilde{h}} 
\def\ft{\tilde{\f}}
\def\gt{\tilde{g}}
\def\pt{\tilde{p}}  
\def\qt{\tilde{q}}  
\def\vt{\tilde{v}}  
\def\nt{\tilde{n}}  
\def\ut{\tilde{u}}  
\def\wt{\tilde{w}}  
\def\zt{\tilde{z}} 
\def\xt{\tilde{x}} 
\def\yt{\tilde{y}} 
\def\Psit{\tilde{\Psi}}
\def\vphit{\tilde{\varphi}}

\def\gamt{\tilde{\gamma}}

\def\Tt{\tilde{T}}

\def\va{{\vec a}}
\def\vk{{\vec k}}
\def\vp{{\vec p}}
\def\vq{{\vec q}}
\def\vx{{\vec x}}
\def\vy{{\vec y}}
\def\vu{{\vec u}}
\def\vv{{\vec v}}

\def\vs{{\vec \sigma}}
\def\vtau{{\vec \tau}}

\newcommand{\ov}[1]{\overrightarrow{#1}}

\def\d{\delta}\def\D{\Delta}\def\ddt{\dot\delta}  

\def\pa{\partial} \def\del{\partial}    

\def\rar{\rightarrow}  
\def\lar{\leftarrow}  
\def\lrar{\leftrightarrow}  

\def\vac{|0\rangle}  
\def\lvac{\langle 0|}  

\def\hlf{\frac{1}{2}}  
\def\ove#1{\frac{1}{#1}}  

\def\CC {\mathbb{C}}
\def\FF {\mathbb{F}}
\def\RR{\mathbb{R}}
\def\NN{\mathbb{N}}  
\def\ZZ{\mathbb{Z}}  
\def\bb#1{{\bf #1}}  
\def\bcomment#1{}  
\def\bfhat#1{{\bf \hat{#1}}}  
\def\VEV#1{\left\langle #1\right\rangle}

\newcommand{\lrbrk}[1]{\left(#1\right)}
\newcommand{\lrsbrk}[1]{\left[#1\right]}
\newcommand{\lrcbrk}[1]{\left\{#1\right\}}
\newcommand{\sfrac}[2]{{\textstyle\frac{#1}{#2}}}

\def\eh{{\hat{e}}}
\def\fh{{\hat{f}}}
\def\lh{{\hat{l}}}
\def\rh{{\hat{r}}}
\def\wh{{\hat{w}}}
\renewcommand{\mh}{{\hat{m}}}

\def \DBI{{\text{DBI}}}
\def\et{{\tilde{\e}}}
\def\w{{\wedge}}
\def\bbV{{\mathbb{V}}}
\def\M{{(\text{M})}}
\def\T{{(\text{T})}}

\def\Hbt{{\tilde{\bar{H}}}}
\def\Fbt{{\tilde{\bar{F}}}}

\def\Pt{{\tilde{P}}}

\def\Fth{\hat{\Ft}}
\def\Hth{\hat{\Ht}}

\numberwithin{equation}{section}

\begin{document}

\title[Double dimensional reduction of M5-brane action in Sen formalism]{Double dimensional reduction of M5-brane action in Sen formalism}


\author[1]{\fnm{Anajak} \sur{Phonchantuek}}\email{anajakp61@nu.ac.th}

\author*[1]{\fnm{Pichet} \sur{Vanichchapongjaroen}}\email{pichetv@nu.ac.th}

\affil[1]{\orgdiv{The Institute for Fundamental Study ``The Tah Poe Academia Institute'', 
		Naresuan University}, \orgaddress{Phitsanulok, \postcode{65000}, \country{Thailand}}}


\abstract{In this paper, we present double dimensional reduction of the complete M5-brane action in the Sen formalism of self-dual form. Although in this formalism the gravity couples to the independent pseudo-$2$-form and pseudo-$3$-form fields in a non-standard and very complicated way, the double dimensional reduction on the complete action can be carried out. This is by dualising some components of the pseudo-$2$-form field and integrating out the pseudo-$3$-form field. We show explicitly that the double dimensional reduction on a circle gives rise to the complete D4-brane or the complete dual D4-brane action depending on what components of the pseudo-$2$-form field are dualised. The duality between the D4-brane and dual D4-brane actions are realised in the viewpoint of the Sen formalism as the swapping of the roles of the components of the pseudo-$2$-form field between those which are dualised and those which remain. We also make a brief outline on how to generalise this to the cases of dimensional reduction on some other spaces as well as how to understand the duality of reduced action from the viewpoint of the Sen formalism. An explicit example for dimensional reduction of the quadratic six-dimensional action on a torus is given and the realisation of its S-duality is also discussed.}

\maketitle

\section{Introduction}\label{sec1}

The constructions of the low-energy effecitive action describing an M5-brane coupled to the background eleven-dimensional supergravity have been known to be non-trivial. This is largely due to the difficulty in the construction of a six-dimensional action for a chiral 2-form field, which is one of the fields in the field content of the M5-brane. The self-duality of its 3-form field strength $H$ makes the straightforward attempt impossible, since $H\w *H = 0$ when imposing $H=*H$ in six dimensions. Therefore, the idea is to first work out the six-dimensional action for a chiral 2-form field then generalise to the complete M5-brane action coupled to background eleven-dimensional supergravity. Various non-trivial approaches which allow this are for example by giving up manifest diffeomorphism invariance \cite{Henneaux:1988gg}, \cite{Perry:1996mk}, \cite{Schwarz:1997mc}, \cite{Aganagic:1997zq}, \cite{Ko:2016cpw}, or by introducing auxiliary fields which have no dynamics as in the PST formalism \cite{Pasti:1997gx}, \cite{Bandos:1997ui}, \cite{Maznytsia:1998xw}, \cite{Maznytsia:1998yr}, \cite{Chen:2010jgb}, \cite{Huang:2011np}, \cite{Ko:2013dka}, \cite{Ko:2017tgo}.
Once the action of the chiral field is constructed, it is usually relatively simpler to extend to the complete M5-brane action.
It is also worth mentioning a recent approach \cite{Mkrtchyan:2019opf}, \cite{Avetisyan:2022zza}, \cite{Evnin:2022kqn} in which further auxiliary fields are introduced to lift the limitation on the auxiliary field in the chiral 2-form action in the PST formalism.

Another recent approach is given in \cite{Sen:2015nph}, \cite{Sen:2019qit} which is motivated by string field theory. This approach, which is called the Sen formalism, in fact applies to self-dual fields in $4p+2$ dimensions. The case relevant to us is therefore $p=1.$ In this case, the Sen formalism gives a new perspective on the $2$-form field with linear self-dual field strength that the $2$-form field is in fact not an independent field. In this approach, the independent fields are a $2$-form field $P$ and a $3$-form field $Q$. Both of these fields are non-standard, for example, they transform in an unusual way under diffeomorphism transformation. Furthermore, the field $Q$ is linear self-dual with respect to flat six-dimensional metric even if the six-dimensional spacetime is curved. It is also required that a certain combination of $Q$ and the curved metric gives rise to a $3$-form field $H$ which is off-shell self-dual with respect to the curved metric and is closed on-shell. So in the spacetime with trivial topology, $H$ is exact on-shell. Therefore, in the Sen formalism, instead of describing 
the $2$-form field with linear self-dual field strength, one describes an exact linear self-dual $3$-form field which is in fact not an independent field.

Studies of various aspects as well as extensions of the Sen formalism are given for example in \cite{Lambert:2019diy}, \cite{Andriolo:2020ykk}, \cite{Lambert:2020scy}, \cite{Gustavsson:2020ugb}, \cite{Vanichchapongjaroen:2020wza}, \cite{Andriolo:2021gen}. Of particular interest to us is the extension \cite{Vanichchapongjaroen:2020wza} to a complete M5-brane action coupled to background eleven-dimensional supergravity. The construction is within the Green-Schwarz formalism in which only the target space has manifest supersymmetry. The constructed action has all the required symmetries for example gauge symmetries, diffeomorphism, and $\k$-symmetry.

Although the M5-brane action in the Sen formalism has been constructed,
further studies are still required in order to better understand the action. One of the important aspects is that 
the independent fields $P$ and $Q$ are expected to couple to gravity in a non-standard and complicated way since these fields transform in an unusual way under diffeomorphism and the action is non-linear. In order to work towards this goal, one may study dimensional reduction.

It turns out that even the dimensional reduction of the original quadratic action in Sen formalism is readily a challenge. Although it is possible to carry out dimensional reduction on various spaces, the studies \cite{Sen:2019qit}, \cite{Lambert:2019diy} are mostly being restricted to the cases in which the uncompact space is flat. So when extending to the M5-brane action, the situation is expected to be even more complicated. Nevertheless, this is exactly the task we focus on in this work.

The aim of this paper is to present double dimensional reduction
of the complete M5-brane action in Sen formalism. In particular, we are going to show explicitly that double dimensional reduction on a circle gives rise to the complete D4-brane action or the complete dual D4-brane action, depending on the processes. It turns out that these two processes are related simply by swapping the components of the psuedo-form $P.$ In fact, the swapping is shown to be related to a duality transformation. We also outline how to generalise this to some other spaces provided that the six-dimensional worldvolume is a Cartesian product and that a chosen set of projection operators on 3-form fields satisfy several simple properties.

This paper is organised as follows. In section \ref{sec:review}, we review the M5-brane action in the Sen formalism. This action is a complete action of M5-brane in the eleven-dimensional supergravity background. Basic properties of this action are reviewed. In particular, we give a brief explanation on how the coupling of this action to gravity is complicated. In section \ref{sec:ddim-circle}, we present double dimensional reduction of the M5-brane action in the Sen formalism on a circle. We first review and slightly modify the approach of \cite{Andriolo:2020ykk} for the dimensional reduction of the Sen quadratic action on a circle. By a further modification, the approach can be applied to the M5-brane action in the Sen formalism whose induced metric is in a general form. The approach involves integrating out some components of the pseudo-form $P,$ while the remaining components are combined with some other fields, through appropriate field redefinitions, to become standard fields.
Depending on the way that the components are chosen, one arrives to either the complete D4-brane or the complete dual D4-brane actions each of which is coupled to type IIA supergravity background. The duality between the D4-brane and the dual D4-brane actions can be realised from the viewpoint of the Sen formalism as coming from swapping of the roles of the components of $P.$ With the insight gained from the analysis, we proceed in section \ref{sec:dimred-otherspaces} to generalise to the cases of dimensional reduction on other spaces. The generalisation looks promising as it gives the reduced action whose physical part describes standard fields. Dualities for reduced actions can also be given from the viewpoint of the Sen formalism. Further checks and generalisations could also be possible which we leave as future works. We give conclusions and discussions in section \ref{sec:conclusion}.
\section{The M5-brane action with self-dual 3-form}\label{sec:review}
In this section, we review the form of the M5-brane action in the Sen formalism \cite{Sen:2015nph}, \cite{Sen:2019qit} which makes use of non-standard self-dual 3-forms in the sense that the self-duality is with respect to flat metric although the theory may couple to the curved metric.

In the Sen formalism of self-dual $(2p+1)$-form in $4p+2$ dimensional spacetime, there are two kinds of Hodge star operators. The first kind is the standard Hodge star operator $*$ which is defined in the standard way, dependent on the spacetime metric. The second kind is denoted $*'$ which is similar to the standard Hodge star operator but is defined with respect to $4p+2$ dimensional Minkowski metric.

From now on, let us focus on the case $p = 1$ which corresponds to self-dual $3$-form in six dimensional spacetime. In this case,
the action takes the form \cite{Sen:2015nph}, \cite{Sen:2019qit}, \cite{Lambert:2019diy},  \cite{Andriolo:2020ykk}
\be\label{Sen-action}
S=\int\lrbrk{\frac{1}{4}dP\w {*'dP}-Q\w dP+\cL_I(Q,g,J)},
\ee
where $P$ is a $2$-form field, $Q$ is a $3$-form field, $g$ is the six dimensional metric, and $J$ is a $3$-form external source.
Note that we have made the rescaling of the fields with respect to those originally given in \cite{Sen:2015nph}, \cite{Sen:2019qit} in order for the action to have the standard scaling. This is closely related to the scaling adopted in \cite{Lambert:2019diy}.
The field $Q$ is $*'$-self-dual, that is $Q = *' Q.$ The variation of the action of $\cL_I$ with respect to $Q$ takes the form
\be
\d_Q\cL_I = \d Q\w R(Q, g, J).
\ee
Therefore, $R$ is $*'$-anti-self-dual. Define
\be
H^J \equiv Q - R + J,
\ee
which should be $*$-self-dual off-shell. That is, for linear theory, $H^J = {*H^J}.$ The equations of motion for $P$ and $Q$ imply that
\be\label{Hs-HJ}
d H^{(s)} = 0,\qquad
d H^J = dJ,
\ee
where
\be
H^{(s)} = Q + \hlf(dP + {*' dP}).
\ee
The kinetic term of the field $P$ in the action \eq{Sen-action} has the wrong sign. So the field $P$ is unphysical. However, this is not harmful as the combination $H^{(s)}$ decouples from the physical field $H^J.$ From eq.\eq{Hs-HJ}, it can be seen that in spacetime with trivial topology, $H^J - J$ is exact on-shell. 

It turns out that when promoting to non-linear theory \cite{Vanichchapongjaroen:2020wza}, the only change in the above discussion is simply that the linear $*$-self-duality becomes the non-linear $*$-self-duality condition
\be\label{HscV}
{*H^J} = \cV(H^J, g).
\ee
In particular, 
a complete M5-brane action in the Green-Schwarz formalism in the form of eq.\eq{Sen-action} reads \cite{Vanichchapongjaroen:2020wza}
\be\label{SM5-mod}
\begin{split}
	S_{M5}=\hlf\int\bigg(&\frac{1}{2}dP\w {*'dP}-2Q\w dP-\frac{1}{12}{*U(F,g)}\\
	&+Q\w R+2C_6+F\w C_3\bigg),
\end{split}
\ee
which describes a six-dimensional worldvolume embedded in the eleven-dimensional background target superspace. For definiteness, let us call the action \eq{SM5-mod} as the Sen M5-brane action. The fields $g, C_3,$ and $C_6$ are induced from the background target superspace.
In the context of the action \eq{Sen-action}, the field $C_3$ is identified as $J$ whereas the field $F$ is identified as $H^J.$ That is
\be
F = Q-R + C_3.
\ee
Note that we have rescaled the action eq.\eq{SM5-mod} with respect to that presented in \cite{Vanichchapongjaroen:2020wza}. This is in order for the scaling to better match with other M5-brane actions presented in the literature. The scaling we made with respect to \cite{Vanichchapongjaroen:2020wza} is $ C_6\to C_6/2, S_{M5}\to S_{M5}/2.$
The explicit form of $\cV$ in eq.\eq{HscV} in the context of M5-brane and the form of $U_{M5}$ in the action \eq{SM5-mod} can be described in index notation. So let us define the convention before coming back to express $\cV$ and $U.$

Let middle Greek alphabets for example $\m,\n,\r$ represent indices
of the coordinates of the six-dimensional worldvolume. The worldvolume coordinates are then labelled as $dx^\m.$
Define $d^6x$ through
\be
dx^{\m_1}\w dx^{\m_2}\w\cdots\w dx^{\m_6}
=d^6x\e^{\m_1\m_2\cdots\m_6},
\ee
where $\e^{\m_1\m_2\cdots\m_6}$ is the Levi-Civita symbol defined such that $\e^{012345} = 1.$
The $*$ operator is defined as
\begin{equation}
	\begin{split}
		*(&d x^{\mu_1}\w\cdots\w dx^{\mu_p})\\
		&=d x^{\mu_{p+1}}\w{\cdots}\w dx^{\mu_6}\frac{(-1)^{p+1}}{(6-p)!\sqrt{-g}}\times\\
		&\qquad g_{\mu_{p+1}\nu_{p+1}}\cdots g_{\mu_6\nu_6}\epsilon^{\nu_{p+1}\cdots \nu_6\mu_1\cdots \mu_p}.
	\end{split}
\end{equation}
Similarly, the $*'$ operator is defined as
\begin{equation}
	\begin{split}
		*'(&d x^{\mu_1}\w\cdots\w dx^{\mu_p})\\
		&=d x^{\mu_{p+1}}\w{\cdots}\w dx^{\mu_6}\frac{(-1)^{p+1}}{(6-p)!}\times\\
		&\qquad\h_{\mu_{p+1}\nu_{p+1}}\cdots \h_{\mu_6\nu_6}\epsilon^{\nu_{p+1}\cdots \nu_6\mu_1\cdots \mu_p}.
	\end{split}
\end{equation}
Differential $p$-forms are expressed in coordinates as
\be
\o_p
=\ove{p!}dx^{\m_1}\w dx^{\m_2}\w\cdots\w dx^{\m_p}\o_{\m_p\cdots\m_2\m_1}.
\ee
Exterior derivative is defined to act from the right
\be
d\o_p
=\ove{p!}dx^{\m_1}\w dx^{\m_2}\w\cdots\w dx^{\m_p}\w dx^\n\pa_\n\o_{\m_p\cdots\m_2\m_1}.
\ee

For M5-brane, the non-linear $*$-self-duality condition \eq{HscV} can be expressed as
\be\label{nonlin-M5}
({*F})_{\m\n\r}
=\lrbrk{-\frac{U}{12}+\frac{24}{U}}F_{\m\n\r} +\frac{6}{U}(F^3)_{\m\n\r},
\ee
where
\be
(F^3)_{\m\n\r} \equiv F_{[\m|\n'\r'}F^{\m'\n'\r'}F_{\m'|\n\r]},
\ee
and $U,$ which also appears in the action \eq{SM5-mod}, is given by
\be\label{Uexpr}
U
=-24\sqrt{1 + \frac{F_{\m\n\r}F^{\m\n\r}}{24}}.
\ee
The indices in eq.\eq{nonlin-M5}-\eq{Uexpr} are raised by $g^{\m\n}.$

The action \eq{SM5-mod} possesses all the required symmetries for example diffeomorphism, gauge symmetries, and kappa-symmetry. It is interesting to note that for each symmetry, transformation rules for $P$ and $Q$ are related by
\be
\d Q = -\hlf(1+*')d\d P,
\ee
which implies that $H^{(s)}$ is invariant under all of the symmetry transformations. This is desirable since $H^{(s)}$ decouples from physical fields.
Its symmetry transformations should be zero in order to not affect the symmetry transformations of physical fields. Focusing on diffeomorphism, the transformations on $P$ and $Q$ under $x^\m\mapsto x^\m + \xi^\m$ are
\be
\d_\xi P = i_\xi(F - C_3),\quad
\d_\xi Q = -\hlf(1+*)di_\xi(F - C_3),
\ee
which are non-standard for $2$-form and $3$-form fields. Due to this, the fields $P$ and $Q$ are called pseudo-forms.

Although the Sen M5-brane action \eq{SM5-mod} is shown to be a complete M5-brane action, further studies are still required in order to better understand this action. Most notably, since $P$ and $Q$ transform in a non-standard way under diffeomorphism, their coupling to gravity are non-standard and is in fact complicated. This complication is readily seen in the quadratic action,
\be\label{quad-action}
S = \int\lrbrk{\frac{1}{4}dP\w {*'dP}-Q\w dP+\hlf Q\w\cM Q},
\ee
which for definiteness, we will call this as the Sen quadratic action.
Here, $\cM$ is a linear operator on $3$-form and is given by\footnote{In principle, there are many possible ways to define the operator $\cM.$ See \cite{Sen:2015nph}, \cite{Sen:2019qit}, \cite{Andriolo:2020ykk} for alternative definitions.}
\be
\cM = (1+**')^{-1}(1-**').
\ee
Part of non-standard coupling to gravity is due to the operator $**'.$ Its appearance through $\cM$ makes this non-standard coupling even more complicated since $\cM$ can in principle be expressed as a power series in $**'$ up to the 19th order.
Indeed, the complication is even further amplified in the Sen M5-brane action as it is non-linear in $Q.$ It is not even clear whether it can be given in the closed form when expressed explicitly in $Q.$

One way to better understand how the Sen M5-brane action is coupled to gravity is to study dimensional reduction, which provides a simpler scenario to study when compared with the Sen M5-brane action itself.

In \cite{Sen:2019qit}, \cite{Andriolo:2020ykk}, dimensional reductions of the Sen quadratic action are carried out on various spaces. Since the action couples to gravity in a complicated way, in each case a special metric is chosen such that the uncompactified space is flat. This makes it possible to carry out the analysis and is sufficient for the purpose of extracting some physics. It was remained to see whether the cases of more general metric can be studied without much difficulties.

In the next section, we will present the study of the double dimensional reduction on a circle of the complete Sen M5-brane action which is coupled to the background eleven-dimensional target superspace. We will demonstrate how to successfully study the full case in which the induced metric takes a general form dependent on the coordinates of reduced worldvolume.
\section{Double dimensional reduction on a circle}\label{sec:ddim-circle}

\subsection{Dimensional reduction of the Sen quadratic action}
Let us review the analysis of dimensional reduction of the Sen quadratic action \eq{quad-action} on a circle. In \cite{Andriolo:2020ykk}, the analysis is given both in the Hamiltonian and Lagrangian formalisms. Let us however only review the analysis of \cite{Andriolo:2020ykk} in the Lagrangian formalism and only follow the key ideas. The exact analysis will be carried out slightly differently especially at the last stage. This results in the interpretation being slightly improved.

Consider the case where the six-dimensional spacetime has the metric of the form
\be\label{getar}
g=
\begin{pmatrix}
	\h_5 & 0\\
	0 & r^2
\end{pmatrix},
\ee
where $\h_5$ is the five-dimensional flat metric,
and $r$ is a positive real number. The coordinate $x^5$ has length dimension and is compactified as $x^5\sim x^5 + l.$
Let every field be independent from the coordinate $x^5$.

Let Roman alphabets for example $a,b,c,i,j,k,m,n$ represent indices of the coordinates of the five-dimensional spacetime.
Denote
\be
\e_{ijkmn}\equiv \e_{ijkmn5},\qquad
\e^{ijkmn}\equiv \e^{ijkmn5}.
\ee
Therefore, $\e_{ijkmn}$ and $\e^{ijkmn}$ are Levi-Civita symbols on the five-dimensional spacetime.

Due to $*'$-self-duality of $Q$ and $*'$-anti-self-duality of $R,$ each of them has only ten independent components. Let us denote indepedent components for each of $Q$ and $R$ as
\be
q_{ab} \equiv Q_{ab5},\qquad
R_{ab} \equiv R_{ab5}.
\ee
Therefore,
\be
\begin{split}
	Q_{abc} &= \hlf\e_{abcij}\h^{im}\h^{jn}q_{mn},\\
	R_{abc} &= -\hlf\e_{abcij}\h^{im}\h^{jn}R_{mn}.
\end{split}
\ee
Let us express $R_{ab}$ in terms of $q_{ab}.$
This can be done by computing
\be
R = \cM Q,
\ee
which gives
\be
R_{ab}
=-\frac{r-1}{r+1}q_{ab}.
\ee
The action \eq{quad-action} then becomes
\be\label{quad-action-dimred-2}
\begin{split}
	S=&\frac{l}{2}\int d^5x\bigg(\frac{3}{4}\pa_a P_{bc}\pa^{[a}P^{bc]} + \pa_a P_{b5}\pa^{[a}P^{b]}{}_5\\
	&\qquad\qquad+2q^{ab}\pa_a P_{b 5}
	+\hlf q_{ab}\pa_m P_{np}\e^{mnpab}\\
	&\qquad\qquad+\frac{(r-1)}{(r+1)}q^{ab}q_{ab}\bigg).
\end{split}
\ee
Next, let us make a replacement
\begin{equation}\label{XdP}
	X_{abc}=3\partial_{[a}P_{bc]}
\end{equation}
on eq.\eq{quad-action-dimred-2} and introduce a Lagrange multiplier $K_m$ which imposes $\pa_{[a}X_{bcd]} = 0.$
The action then becomes
\be
\begin{split}
	S=&\frac{l}{2}\int d^5x\bigg(\frac{1}{12}X_{abc}X^{abc}+\frac{1}{6}q_{ab}\epsilon^{abijk}X_{ijk}\\
	&\qquad\qquad+\frac{1}{6}X_{abc}\partial_mK_n\epsilon^{abcmn}+\partial_aP_{b5}\partial^{[a}P^{b]}{}_{5}\\
	&\qquad\qquad+2q^{ab}\partial_aP_{b5}+\frac{(r-1)}{(r+1)}q^{ab}q_{ab} \bigg).
\end{split}
\ee
By integrating out $X$ followed by $q,$ we obtain
\be
\begin{split}
	S
	&=\frac{l}{2}\int d^5 x\bigg( \frac{r-1}{2r}\pa_a K_b \pa^{[a} K^{b]}\\
	&\qquad\qquad-\frac{r+1}{r}\pa_a K_b \pa^{[a} P^{b]5}\\
	&\qquad\qquad+\frac{r-1}{2r}\pa_a P_{b5} \pa^{[a} P^{b]}{}_5\bigg).
\end{split}
\ee
We may view the quantity inside the bracket as a quadratic form
whose matrix
\be
\ove{2r}
\begin{pmatrix}
	r-1 & -(r+1)\\
	-(r+1) & r-1
\end{pmatrix}
\ee
has eigenvalues $-1/r,1$ and the corresponding eigenvectors $(1,1)^T, (1,-1)^T.$
Therefore, let us define
\begin{equation}
	f_{ab}=l(\partial_{[a} K_{b]}+\partial_{[a} P_{b]}),\quad
	f^{(s)}_{ab}=l(\partial_{[a} K_{b]}-\partial_{[a} P_{b]})
\end{equation}
where the factor $l$ is introduced so that $f$ and $f^{(s)}$ have mass dimension two.
This corresponds to the field redefinition
\be\label{A-As}
A_a = \frac{l}{2} (K_a + P_a),\quad
A^{(s)}_a = \frac{l}{2} (K_a - P_a).
\ee
So $f_{ab}$ and $f^{(s)}_{ab}$ are field strengths of $A_a$ and $A^{(s)}_a,$ respectively.
With the field redefinitions, the action becomes
\be
\begin{split}
	S=&\int d^5x\lrbrk{-\frac{1}{4rl}f_{ab}f^{ab}+\frac{1}{4l}f^{(s)}_{ab}(f^{(s)})^{ab}},
\end{split} 
\ee
which describes two uncoupled Maxwell fields with opposite signs of kinetic terms. For any value of $r\in\RR^+,$ the field $A_a$ has the correct sign of the kinetic term with the $1/r$ scaling as required by conformal symmetry \cite{Witten:2009at}. On the other hand, the field $A^{(s)}_a$ has the wrong sign of the kinetic term.

Note that the difference between the analysis in \cite{Andriolo:2020ykk} and ours is just only in the field redefinition which we have made in eq.\eq{A-As}. The counterpart presented in \cite{Andriolo:2020ykk} leads to the conclusion that the kinetic term of the standard Maxwell field scales with $1/r$ only in the limit $r\to 0.$
On the other hand, the field redefinition \eq{A-As} leads to the conclusion that the kinetic term of the standard Maxwell field scales with $1/r$ for any value of $r\in\RR^+,$ not only in the $r\to 0$ limit. The $1/r$ scaling upon dimensional reduction on a circle is also in agreement with the result from the PST formalism. As a simple check, consider the quadratic version of the M5-brane action in dual formulation. By considering the six-dimensional spacetime with metric \eq{getar}, identifying $x^5$ as a coordinate on the circle, and keeping only the zero Fourier mode,
the action presented in eq.(3.19) of the reference \cite{Ko:2016cpw} reduces to the five-dimensional Maxwell theory with the scaling $1/r.$ The exact agreement with the requirement from conformal symmetry and the relevant result from the PST formalism further confirms that the Sen quadratic action is a quadratic action for an M5-brane.

In the next subsections, we will proceed to extend the idea of this subsection to analyse double dimensional reductions of the Sen M5-brane action on a circle. It will be seen that although the Sen M5-brane action couples to gravity in a complicated way, the analysis based on slight modifications of the steps given in this subsection turns out to apply also for the Sen M5-brane action.

\subsection{The complete D4-brane action from the Sen M5-brane action}
In the previous subsection, we have reviewed the dimensional reduction of the Sen quadratic action to give a five-dimensional Maxwell theory.
We worked with the case where the metric is given by eq.\eq{getar}, in which the six-dimensional worldvolume is a cartesian product of a flat five-dimensional spacetime with a circle with coordinate $x^5$.

It is possible, after slightly modifying the strategy, to extend the analysis to the case of the Sen M5-brane action which is a non-linear action with source terms.
In the previous subsection, we have seen that the dimensional reduction
is obtained by first expressing $R$ in terms of $Q,$ dualising some components of $P,$ then using Euler-Lagrange equations to eliminate $Q.$ Then after field redefinition, we are left with the action which describes an unphysical field decoupled with Maxwell field in five dimensions.

In this section, since we are considering a
non-linear action with source terms, expressing $R$ directly in terms of $Q$ would be too complicated if at all possible. 
As to be given in details below, it turns out that the modification of the strategy is simply to switch some steps. This would allow us to successfully dimensionally reduce the Sen M5-brane action to the complete action which describes a D4-brane in type IIA supergravity background uncoupled with an unphysical field.

Let us consider a double dimensional reduction of the Sen M5-brane action \eq{SM5-mod} by letting
the background target space be compactificed such that $X^{10}$ is on a circle
and let the $x^5$ coordinate of M5-brane worldvolume wraps around $X^{10}.$
The length of $x^5$ is given by
\be
\int dx^5 = 2\p g_s l_s.
\ee
So if the tension $T_{M5}$ of the M5-brane action \eq{SM5-mod} is introduced back, which is simply by making a scaling
$S_{M5}\mapsto T_{M5}S_{M5}$, then $T_{M5} 2\p g_s l_s = T_{D4}$ as expected.
So from now on, let us set for simplicity
\be
\int dx^5 = 1,\qquad
T_{M5} = 1 = T_{D4}.
\ee

Under the double dimensional reduction, the induced metric on the worldvolume is given by
\be
g_{\m\n}=
\begin{pmatrix}
	e^{-2\f/3}\g_{ab} + e^{4\f/3}C_a C_b & e^{4\f/3}C_a\\
	e^{4\f/3} C_b & e^{4\f/3}
\end{pmatrix}.
\ee
The fields $\f$ and $C_a$ are considered as pullbacks of the background superfield
to five-dimensional worldvolume.
Let us next consider how the pullbacks $C_3$ and $C_6$ of the corresponding target space superfields are dimensionally reduced.
The components of $C_3$ are separated via
\be
C_3 = \hlf dx^a\w dx^b\w dx^5 C_{ba5} + \ove{3!}dx^i\w dx^j\w dx^k C_{kij}.
\ee
It is natural to denote the following:
\be
b_{ab} \equiv C_{ab5},\qquad
\tilde{C}^{ab} \equiv \ove{3!}\e^{abijk}C_{ijk}.
\ee
Note by definition that $\tilde{C}^{ab}$ is the dualisation of $C_{ijk}$
with respect to five-dimensional flat metric $\h.$
So it would be convenient to define raising and lowering of the indices of $B$ and $\tilde{C}$ by using five-dimensional flat metric.
As for $C_6,$ it can be expressed as
\be
C_6 = -C_5\w dx^5.
\ee
So in summary, for the reduced theory, the pullbacks of the target space superfields of type IIA supergravity to the five-dimensional worldvolume theory are $\g_{ab}, b_2, C_1, C_3, C_5.$

The worldvolume action then reads
\be\label{S-pred-D4}
\begin{split}
	S
	&=\hlf\int d^5x\bigg(\frac{3}{4}\pa_a P_{bc}\pa^{[a}P^{bc]}
	+ \pa_a P_{b5}\pa^{[a}P^{b]}{}_5\\
	&\qquad\quad
	+2q^{ab}\pa_a P_{b 5}
	+\hlf q_{ab}\pa_m P_{np}\e^{mnpab}\\
	&\qquad\quad
	+ \frac{1}{12}\sqrt{-g}U + [q(R+b)]\\
	&\qquad\quad - \hlf[(q-R)(b-\Ct)]\bigg)
	-\int C_5.
\end{split}
\ee
Since, unlike the case discussed in the previous subsection, the five-dimensional part of the worldvolume is not flat.
So one has to be more careful when defining trace as there are two kinds of metric being used here to raise and lower indices.
Let us revisit the definition of trace.
From now on, it will be understood that components of matrices appearing within the trace are of the form $(\cdot)^a{}_b.$
For example,
\be
[MN]
=M^{a}{}_b N^{b}{}_{a}.
\ee
For a given five-dimensional matrix $M,$ we will specify whether its components are raised by five-dimensional flat metric $\h$ or five-dimensional curved metric $\g.$
The components of matrices $q, R, B, \Ct$ which we encounter so far are raised and lowered by $\h.$ We will also later define other five-dimensional matrices and will specify the metric which are used to raise or lower their components.

If we were to follow the same strategy as that of the previous section, we would have to eliminate $R$ from eq.\eq{S-pred-D4} by expressing it in terms of $Q, g, C_3.$ Here, however, let us instead keep $R$ explicit at this stage and simply proceed to dualise $P_{ab}.$ By defining $X_{abc}$ as in eq.\eq{XdP} and introducing the Lagrange multiplier term, we obtain
\be\label{S-pred-D4-2}
\begin{split}
	S
	&=\hlf\int d^5x\bigg(\frac{1}{12}X_{abc}X^{abc} + \pa_a P_{b5}\pa^{[a}P^{b]}{}_5\\
	&\qquad\quad
	+2q^{ab}\pa_a P_{b 5}
	+\ove{6} q_{ab}X_{mnp}\e^{mnpab}\\
	&\qquad\quad
	+\ove{3!}X_{abc}\pa_m K_n\e^{abcmn}
	+ \frac{1}{12}\sqrt{-g}U\\
	&\qquad\quad
	+ [q(R+b)] - \hlf[(q-R)(b-\Ct)]
	\bigg)\\
	&\qquad-\int C_5.
\end{split}
\ee
Integrating out $X$ gives
\be\label{S-pred-D4-3}
\begin{split}
	S
	&=\hlf\int d^5x\bigg(-\lrsbrk{(\pa K + \pa P + q)^2}
	+2\lrsbrk{\pa K\pa P}\\
	&\qquad
	+\frac{1}{12}\sqrt{-g}U + [q(R+b)]\\
	&\qquad- \hlf[(q-R)(b-\Ct)]
	\bigg)
	-\int C_5,
\end{split}
\ee
where $\pa K$ and $\pa P$ are matrices with components $(\pa K)_{mn}\equiv \pa_{[m} K_{n]}$ and $(\pa P)_{mn}\equiv\pa_{[m}P_{n]5},$ respectively. The components of $\pa K$ and $\pa P$ are raised and lowered by $\h.$
Let us define
\be\label{field-redef-1}
\psi\equiv\pa K + \pa P,\qquad
\psi^{(s)}\equiv\pa K - \pa P,
\ee
whose components are also raised and lowered by $\h.$
We may view the above definition as the field redefinitions
\be\label{field-redef-2}
K_m+P_m\equiv -\frac{A_m}{2},\qquad
K_m-P_m\equiv \frac{A^{(s)}_m}{2}.
\ee
So $-\psi$ and $\psi^{(s)}$ are field strengths of $A$ and $A^{(s)}.$
The action becomes
\be\label{S-pred-D4-4}
\begin{split}
	S
	&=\hlf\int d^5x\bigg(-\lrsbrk{(\psi + q)^2}
	+\hlf[\psi^2] - \hlf[(\psi^{(s)})^2]\\
	&\qquad\quad
	+ \frac{1}{12}\sqrt{-g}U + [q(R+b)]\\
	&\qquad\quad - \hlf[(q-R)(b-\Ct)]
	\bigg)
	-\int C_5.
\end{split}
\ee
Euler-Lagrange equation for $q$ gives
\be\label{Rfq}
R = \psi+q.
\ee
By using eq.\eq{Rfq} to eliminate $R$ from the action, we obtain
\be\label{S-D4-elim-R}
\begin{split}
	S
	&=\int d^5 x\bigg(-\ove{4}[(\psi^{(s)})^2]
	-\ove{4}[\psi^2]
	+ \hlf[q(-\psi+b)]\\
	&\qquad\qquad\qquad+ \frac{1}{24}\sqrt{-g}U
	+\ove{4}[\psi b]
	-\ove{4}[\psi\tilde C]\bigg)\\
	&\qquad\qquad\qquad-\int C_5.
\end{split}
\ee

It is possible to simplify the expression for $U.$
Let us define matrices $F$ and $\Ft$ whose components are
\be
F^a{}_{b} \equiv e^{-2\f/3}F^a{}_{b5},\quad
\Ft^a{}_{b} \equiv e^{-2\f/3}\Ft^a{}_{b5}.
\ee
It will be understood that indices of $F_{ab}$ and $\Ft_{ab}$ are raised and lowered by $\g.$
With this convention, we have
\be\label{FF}
F_{\m\n\r}F^{\m\n\r}
=-3[F^2] + 3[\Ft^2].
\ee

Since twenty components of $F_{\m\n\r}$ are related to each other by the non-linear self-duality condition, only ten independent components are required. For our case, we need to eliminate $\Ft_{ab5}.$ The relevant non-linear self-duality condition is
\be\label{Ft-as-F}
\Ft_{ab}
=\frac{(1-y_1)F_{ab} + (F^3)_{ab}}{\sqrt{1-y_1 + y_1^2 - y_2}},
\ee
where
\be
y_1\equiv\hlf [F^2],\qquad
y_2\equiv\ove{4} [F^4].
\ee
Therefore, after substituting into eq.\eq{FF} and then working out $U,$
we obtain
\be
U=
-12
\frac{2- y_1}{\sqrt{1-y_1 + \hlf y_1^2 - y_2}}. 
\ee

Let us eliminate $q.$ By considering the $ijk$ component of $H = Q-R$
along with eq.\eq{Rfq}, we obtain
\be\label{q-D4-final}
\begin{split}
	q^{ab}
	&= \hlf\bigg(\sqrt{-g}\Ft^{ab} + \Ct^{ab}
	- \hlf\e^{abijk}F_{ij}C_k\\
	&\qquad\qquad + \h^{ac}\h^{bd}F_{cd}-b^{ab}\bigg).
\end{split}
\ee
This completely determine $Q$ in terms of other fields due to eq.\eq{Ft-as-F} and
\be
F_{ab} = -\psi_{ab} + b_{ab},
\ee
where, as defined in eq.\eq{field-redef-1}-\eq{field-redef-2}, $\psi_{ab}$ is a field strength of a 1-form field.

So by substituting eq.\eq{q-D4-final} into eq.\eq{S-D4-elim-R}, the action becomes
\be\label{D4-action}
\begin{split}
	S
	&=\int d^5 x\bigg(\ove{4}\psi^{(s)}_{ab}\psi^{(s)}_{ij}\h^{ai}\h^{bj}\\
	&\qquad\qquad\quad-e^{-\f}\sqrt{-\det(\g_{ab} + F_{ab})}\bigg)\\
	&\qquad-\int\lrbrk{C_1+C_3+C'_5}\w e^F,
\end{split}
\ee
where
\be
F_{ab} = \pa_a A_b - \pa_b A_a + b_{ab},
\ee
\be
C'_5
=C_5 - \hlf B\w C_3,
\ee
\be
\psi^{(s)}_{ab}
=\pa_a A^{(s)}_b - \pa_b A^{(s)}_a.
\ee
The action \eq{D4-action} describes
the complete D4-brane \cite{Aganagic:1997zk} in type IIA supergravity background decoupled with unphysical field.

At this point, let us compare with the result from the PST formalism. It can be seen from the discussion in the literature that double dimensional reduction of the M5-brane action in the original PST action \cite{Aganagic:1997zq}, \cite{Pasti:1997gx}, \cite{Bandos:1997ui} on a circle is given by the dual D4-brane action, which can be realised as a duality transformation \cite{Aganagic:1997zk} of the D4-brane action. Alternatively, if one starts with the dual formulation of the M5-brane action \cite{Ko:2016cpw}, \cite{Ko:2017tgo}, then it can be shown that double dimensional reduction on a circle directly gives rise to the D4-brane action.
As for the case of Sen formalism, we have seen in this subsection that essentially by dualising some components of $P,$ we obtain as a result of double dimensional reduction of the complete M5-brane action on a circle, the D4-brane action. Therefore, it seems that the original PST M5-brane action naturally gives rise, after double dimensional reduction on a circle, to dual D4-brane action. On the other hand, the Sen M5-brane action seems to naturally give rise to D4-brane action. In fact, as will be demonstrated in the next subsection, it can be seen that the Sen M5-brane action also naturally give rise to dual D4-brane action.

\subsection{Dual D4-brane action from Sen M5-brane action}

We have learned from the previous section that the key steps of double dimensional reduction of the M5-brane action in Sen formalism to the D4-brane action
are essentially to dualise components $P_{ab}$ of $P$ and then eliminate $Q.$ The five-dimensional DBI gauge field $A$ and unphysical field $A^{(s)}$ then arise as linear combinations of the Lagrange multiplier and the remaining components $P_{a5}$ of $P.$

In fact, by slightly modifying the procedure,
it is possible to obtain the dual D4-brane action
by dimensionally reducing Sen M5-brane action.
The modification is simply to switch the role of the components of $P$
so that we dualise components $P_{a5}$ and as to be seen at the end of the process that components of $P_{ab}$ along with the Lagrange multiplier will be linearly combined to form gauge field and unphysical field.

Let us show this in more details.
We follow the same steps as those of the previous section up to eq.\eq{S-pred-D4}. Then, by defining
\be
Y_{ab}
=2\pa_{[a}P_{b]5}
\ee
and introducing Lagrange multiplier term, we obtain
\be\label{S-pred-dualD4-1}
\begin{split}
	\St
	&=\hlf\int d^5x\bigg(\frac{3}{4}\pa_a P_{bc}\pa^{[a}P^{bc]}
	+\ove{4}Y_{ab}Y^{ab}
	+q^{ab}Y_{ab}\\
	&\qquad\qquad+\hlf q_{ab}\pa_m P_{np}\e^{mnpab}
	+\ove{4}Y_{ab}\pa_m K_{np}\e^{abmnp}\\
	&\qquad\qquad
	+ \frac{1}{12}\sqrt{-g}U + [q(R+\Ct)]\\
	&\qquad\qquad + \hlf[(q+R)(b-\Ct)]
	\bigg)
	-\int C_5.
\end{split}
\ee
Integrating out $Y$ gives
\be\label{S-pred-dualD4-2}
\begin{split}
	\St
	&=\hlf\int d^5x\bigg(\lrsbrk{(\widetilde{\pa K} - \widetilde{\pa P} + q)^2}+2\lrsbrk{\widetilde{\pa K}\widetilde{\pa P}}\\
	&\qquad\quad
	+ \frac{1}{12}\sqrt{-g}U + [q(R+\Ct)]\\
	&\qquad\quad + \hlf[(q+R)(b-\Ct)]
	\bigg)
	-\int C_5,
\end{split}
\ee
where $\widetilde{\pa K}$ and $\widetilde{\pa P}$ are matrices with components $\widetilde{\pa K}^{mn}\equiv\e^{mnijk}\pa_i K_{jk}/4$ and  $\widetilde{\pa P}^{mn}\equiv\e^{mnijk}\pa_i P_{jk}/4,$ respectively. Indices for matrices $\widetilde{\pa K}, \widetilde{\pa P}$ are raised and lowered by the five-dimensional flat metric $\h.$
Let us define
\be\label{fredefn-chi}
\chi\equiv\widetilde{\pa P} - \widetilde{\pa K},\qquad
\chi^{(s)}\equiv\widetilde{\pa P} + \widetilde{\pa K},
\ee
which can be viewed as the field redefinitions
\be
P_{ij}-K_{ij}\equiv 2B_{ij},\qquad
P_{ij}+K_{ij}\equiv 2B^{(s)}_{ij}.
\ee
So $\chi$ and $\chi^{(s)}$ are dual of field strengths of $B$ and $B^{(s)}$
with respect to five-dimensional flat metric.
The action then becomes
\be\label{S-pred-dualD4-3}
\begin{split}
	\St
	&=\hlf\int d^5x\bigg(\lrsbrk{(-\chi + q)^2}
	-\hlf[\chi^2] + \hlf[(\chi^{(s)})^2]\\
	&\qquad\quad
	+ \frac{1}{12}\sqrt{-g}U + [q(R+\Ct)]\\
	&\qquad\quad + \hlf[(q+R)(b-\Ct)]
	)
	-\int C_5.
\end{split}
\ee
Variation with respect to $q$ gives
\be\label{Rfq-dual}
R = \chi-q.
\ee
Eliminating $R$ from the action gives
\be\label{S-pred-dualD4-4}
\begin{split}
	\St
	&=\int d^5x\bigg(\ove{4}[(\chi^{(s)})^2]
	+\ove{4}[\chi^2]
	+ \hlf[q(-\chi+\Ct)]\\
	&\qquad\qquad\qquad+ \frac{1}{24}\sqrt{-g}U
	+\ove{4}[\chi b]
	-\ove{4}[\chi\tilde C]\bigg)\\
	&\qquad-\int C_5.
\end{split}
\ee

The next step is to reexpress $U.$
For this, let us define
\be
\Fh^{a}{}_b \equiv \frac{F^{a}{}_b{}^5}{\sqrt{g^{55}}},\qquad
\Fth^{a}{}_b \equiv \frac{\Ft^{a}{}_b{}^5}{\sqrt{g^{55}}}.
\ee
Let indices of 
$\Fh^{ab}$ and $\Fth^{ab}$ be lowered by $\g_{ab}$ whereas
indices of $F^{\m\n\r}$ and $\Ft^{\m\n\r}$ are lowered by $g_{\m\n}.$
We obtain
\be
\begin{split}
	F_{\m\n\r}F^{\m\n\r}
	&=-3 [\Fh^2] + 3 [\Fth^2].
\end{split}
\ee
We need to eliminate $\Fh_{ab}$ using nonlinear self-duality condition
\be\label{F-as-Ft}
\Fh^a{}_b
=\frac{(1+z_1)\Fth^a{}_b-(\Fth^3)^a{}_b}{\sqrt{1+z_1 + \hlf z_1^2 - z_2}},
\ee
where
\be
z_1\equiv\hlf [\Fth^2],\qquad
z_2\equiv\ove{4} [\Fth^4].
\ee
This gives
\be
U = -\frac{12(2+z_1)}{\sqrt{1+z_1 + \hlf z_1^2 - z_2}}.
\ee

Let us eliminate $q$ from the action \eq{S-pred-dualD4-3}.
By considering the $ab5$ component of $H = Q-R,$ we obtain
\be
\begin{split}
	&q_{ab} - R_{ab} + b_{ab}\\
	&=\ove{\sqrt{g^{55}}}\lrbrk{g_{ac}\Fh^c{}_b
		-\hlf\sqrt{-g}\e_{abcmn}g^{c5}\Fth^m{}_p g^{pn}}.
\end{split}
\ee
So along with eq.\eq{Rfq-dual},
we obtain
\be
\begin{split}
	q_{ab}
	&= \hlf\bigg(\frac{g_{ac}}{\sqrt{g^{55}}}\Fh^c{}_b
	-\hlf\frac{\sqrt{-g}}{\sqrt{g^{55}}}\e_{abcmn}g^{c5}\Fth^m{}_p g^{pn}\\
	&\qquad- b_{ab}+\sqrt{-g}\sqrt{g^{55}}\Fth^{m}{}_p\gt^{pn}\h_{ma}\h_{nb}
	+\Ct_{ab}
	\bigg),
\end{split}
\ee

which completely determine $Q$ in terms of other fields
due to eq.\eq{F-as-Ft}.
Here
\be
\gt^{ab}\equiv g^{ab} - \frac{g^{a5}g^{b5}}{g^{55}}.
\ee
We also have
\be\label{chi-Ft-C}
\chi^{ab}
=\sqrt{-g}\sqrt{g^{55}}\Fth^{a}{}_c\gt^{cb}
+\Ct^{ab}.
\ee
Let us define
\be\label{W-Ft}
\begin{split}
	W^{ab}
	&\equiv\sqrt{-g}\sqrt{g^{55}}\Fth^{a}{}_d\gt^{db}\\
	&=-\ove{3!}\e^{abijk}F_{ijk},
\end{split}
\ee
which is the dualisation of $F_{ijk}$ with respect to five-dimensional flat metric $\h.$ The components of $W$ are raised and lowered by $\h.$

So the action becomes
\be\label{S-dualD4}
\begin{split}
	\St
	&=\int d^5 x\bigg(\frac{3}{4}\pa_{[a}b^{(s)}_{bc]}\pa_{[i}b^{(s)}_{jk]}\h^{ai}\h^{bj}\h^{ck}\\
	&\qquad\qquad-e^{-\f}\sqrt{-\g}\sqrt{1+z_1 + \hlf z_1^2 - z_2}\\
	&\qquad\qquad+\hlf[Wb] + \ove{4}[\Ct b]\\
	&\qquad\qquad-\ove{8}\ove{g^{55}}\e_{abcmn}g^{c5}W^{mn}W^{ab}
	\bigg)-\int C_5\\
	&=\int d^5 x\bigg(\frac{3}{4}\pa_{[a}b^{(s)}_{bc]}\pa_{[i}b^{(s)}_{jk]}\h^{ai}\h^{bj}\h^{ck}\\ 
	&\qquad\qquad
	-e^{-\f}\sqrt{-G}\\
	&\qquad\qquad-\ove{24}\frac{\g^{cd}C_d\e^{abmnp}F_{mnp}F_{abc}}{e^{-4\f/3}+e^{2\f/3} C_i\g^{ij}C_j}\bigg)\\
	&\qquad
	+\int (b\w(F-C_3)-C'_5),
\end{split}
\ee
where
\be
G_{ab}\equiv\g_{ab}-\frac{i}{6}e^{\f}\frac{\g_{ac}\e^{cdijk}F_{ijk}(\g_{db}+e^{2\f}C_dC_b)}{\sqrt{-\g}\sqrt{1+e^{2\f} C_m\g^{mn}C_n}},
\ee
and, due to eq.\eq{fredefn-chi}, \eq{chi-Ft-C}-\eq{W-Ft},
\be
F_{ijk}
=\pa_i B_{jk} + \pa_j B_{ki} + \pa_k B_{ij} +C_{ijk}.
\ee
The action \eq{S-dualD4} describes the complete dual D4-brane action \cite{Aganagic:1997zk} decoupled with unphysical field.

\subsection{Duality from the viewpoint of Sen M5-brane action}\label{subsec:dual-D4-dD4}
It is well known that the D4-brane part of the action \eq{D4-action} and the dual D4-brane part of the action \eq{S-dualD4} are related by duality transformation. This can be shown \cite{Aganagic:1997zk} by considering the dualisation of $A_a$ in the D4-brane action, which then gives rise to 
the dual D4-brane action. More explicitly, one replaces $2\pa_{[a}A_{b]}$ in the D4-brane action by $\psi_{ab},$ treats the latter as an independent field, introduces a $3-$form Lagrange multiplier $\chi$ which puts $\psi_{ab} = 2\pa_{[a}A_{b]},$ then finally integrates out $A_a$ and $\psi_{ab}.$ This process gives rise to the dual D4-brane action, in which the Lagrange multiplier $\chi$ becomes the field strength of the $2$-form field $B.$

It can be shown from the viewpoint of dimensionally reduced Sen M5-brane action that the action \eq{D4-action} and \eq{S-dualD4} are dual to each other. The prove can be performed quite simply by noting that the action \eq{D4-action} is equivalent to the action \eq{S-pred-D4-3}, whereas the action \eq{S-dualD4} is equivalent to the action \eq{S-pred-dualD4-2}.
Let us replace $(\pa K)_{ab}$ and $(\pa P)_{ab}$ in the action \eq{S-pred-D4-3} by $\F^{(1)}_{ab}$ and $\F^{(2)}_{ab},$ respectively. Let us treat $\F^{(1)}_{ab}$ and $\F^{(2)}_{ab}$ as independent fields and impose Lagrange multipliers $\Th^{(1)}_{ab}, \Th^{(2)}_{ab}$ which impose 
$\pa_{[a}\F^{(1)}_{bc]}=0=\pa_{[a}\F^{(2)}_{bc]}.$ So the action \eq{S-pred-D4-3} becomes

\be\label{S-pred-D4-3-alt}
\begin{split}
	S
	&=\int d^5x\bigg(-\hlf\lrsbrk{(\F^{(1)} + \F^{(2)} + q)^2}
	+\lrsbrk{\F^{(1)}\F^{(2)}}\\
	&\qquad\qquad
	+\ove{4}\e^{abcij}(\pa_a\Th^{(1)}_{bc}\F^{(1)}_{ij}+\pa_a\Th^{(2)}_{bc}\F^{(2)}_{ij})\\
	&\qquad\qquad+\frac{1}{24}\sqrt{-g}U+ \hlf[q(R+b)]\\
	&\qquad\qquad- \ove{4}[(q-R)(b-\Ct)]\bigg)
	-\int C_5.
\end{split}
\ee
Euler-Lagrange equations for $\Th^{(1)}_{ab}$ and $\Th^{(2)}_{ab}$ are $\pa_{[i}\F^{(1)}_{jk]} = 0 = \pa_{[i}\F^{(2)}_{jk]},$ which if the five-dimensional spacetime is topologically trivial in the sense that all closed forms are exact, then we can write $\F^{(1)}_{ij} = \pa_{[i}K_{j]}, \F^{(2)}_{ij} = \pa_{[i}P_{j]},$ which upon substitution into eq.\eq{S-pred-D4-3-alt}, we reobtain eq.\eq{S-pred-D4-3} as expected. Let us however consider Euler-Lagrange equations for $\F^{(1)}_{ab}$ and $\F^{(2)}_{ab}$ which imply
\be
(\F^{(1)})^{ab}
=-q^{ab} - \ove{4}\e^{ijkab}\pa_i\Th_{jk}^{(1)},
\ee
\be
(\F^{(2)})^{ab}
=-q^{ab} - \ove{4}\e^{ijkab}\pa_i\Th_{jk}^{(2)}.
\ee
Substituting these results into eq.\eq{S-pred-D4-3-alt}, keeping in mind that indices for $q, \F^{(1)}, \F^{(2)}$ are raised and lowered by $\h,$ and denoting $\Th_{ij}^{(1)} = -P_{ij}$ and $\Th_{ij}^{(2)} = K_{ij},$ we obtain eq.\eq{S-pred-dualD4-2}. This finishes the proof in the viewpoint of Sen formalism that the D4-brane action and the dual D4-brane action are dual to each other.
\section{Towards dimensional reduction of the Sen M5-brane action on other spaces}\label{sec:dimred-otherspaces}

In \cite{Sen:2019qit}, \cite{Andriolo:2020ykk}, analyses of dimensional reductions of the 6d Sen quadratic action are given. The examples of the spaces are a circle, a torus, $K3,$ and a non-compact Riemann surface. Furthermore, in section \ref{sec:ddim-circle} we have demonstrated that when non-linearising to the Sen M5-brane action, the dimensional reduction on a circle can naturally be given. In particular, it is natural to give, based on the procedure of double dimensional reduction, either D4-brane action or dual D4-brane action.
Note that these results suggest a remarkable feature of the Sen formalism that although its quadartic action couples to gravity in a complicated way, dimensional reduction on various cases of spaces can naturally be performed.

In contrast, in the PST formalism, although dimensional reduction can in principle be performed on any spaces, it is more natural to use specific action for dimensional reduction on a specific space.
For example, it is natural to perform dimensional reduction of the original PST action \cite{Aganagic:1997zq}, \cite{Pasti:1997gx}, \cite{Bandos:1997ui} and its dual version \cite{Ko:2016cpw}, \cite{Ko:2017tgo} on a circle giving rise respectively to D4-brane and dual D4-brane actions, whereas it is natural to perform dimensional reduction of the $3+3$ M5-brane action \cite{Ko:2013dka} on $T^3$ to give an M2-brane action.

In this section, we will work towards generalising the analysis in section \ref{sec:ddim-circle} so that it applies also to dimensional reductions on other spaces. We will only focus on special cases which contain some particular ways of dimensional reductions. We will also only give an outline of the analysis. The explicit details will be studied as future works. We expect that the analysis in this section could be served as a base for further generalisations, for example, double dimensional reduction of the complete M5-brane action on $T^{5-p}$ to give D$p$-brane with $p<4$ (the case $p=4$ has been given in the previous sections).

\subsection{An outline of the analysis}\label{subsec:dimred}

Let us consider dimensional reduction of the Sen M5-brane action \eq{SM5-mod}
from a $6$ dimensional spacetime $\cM_6 = \cD_D \times \cC_{6-D}$ with coordinates\footnote{In this section, middle Greek alphabets for example $\m, \n, \r$ are used as indices of coordinates on $\cM_6.$ Roman alphabets for example $a,b,c,i,j,k,m,n$ are used as indices of coordinates on $\cD_D.$
	Beginning Greek alphabets for example $\a,\b$ are used as indices of coordinates on $\cC_{6-D}$.} $x^\m, \m = 0,1,2,\cdots,5$ to a $D$ dimensional topologically trivial spacetime $\cD_D$ with coordinates $x^a, a = 0,1,\cdots,D-1.$
Let fields be independent from the coordinates $x^\a, \a=D,D+1,\cdots,5$ of the compact space $\cC_{6-D}.$
Denote $d$ and $\dh$ to be the exterior derivative on $\cM_6$ and $\cD_D,$ respectively.

On $\cM_6,$ let us denote a 3-form coordinate basis in six dimensions as $\cB = \{dx^\m\w dx^\n\w dx^\r\ |\m,\n,\r\in\{0,1,2,\cdots,5\}\}.$
Let us pick $2n$ projection operators $\cP_I,$ $I = 1,2,\cdots,2n$ such that
\be\label{cPIconst}
\begin{split}
\cP_I(&dx^\m\w dx^\n\w dx^\r)\\
&=
\begin{cases}
	dx^\m\w dx^\n\w dx^\r & \textrm{if }dx^\m\w dx^\n\w dx^\r\in\cB_I,\\
	0 & \textrm{if }dx^\m\w dx^\n\w dx^\r\not\in\cB_I,
\end{cases}
\end{split}
\ee
where
$\cB_1, \cB_2, \cdots, \cB_{2n}$ are mutually exclusive subsets of $\cB = \cup_{I=1}^{2n}\cB_I.$
Therefore, the projection operators satisfy
\be
\sum_{I=1}^{2n} \cP_I = \mathbbm{1},
\ee
\be\label{spPsp}
*'\cP_I*' = Y_{IJ}\cP_J,
\ee
where $Y_{IJ}$ are constants.
Let us also impose a condition that the projection operators $\cP_I$ satisfy the property that for any $2$-form $A$ which is independent from coordinates of $\cC_{6-D},$ there exists a $2$-form $A_I$ such that
\be\label{cPIdA}
\cP_I d A = \dh A_I.
\ee
Since $A$ is independent from $x^\a,$ it can easily be seen that $A_I$ are also independent from $x^\a.$
Let us require furthermore that the first $n$ projection operators $\cP_{\hat{I}}, \hat{I}=1,2,\cdots,n,$
are related to the last $n$ projection operators
$\cP_{\bar{I}}, \bar{I}=n+1,n+2,\cdots,2n,$
by
\be
*'\cP_{\hat{I}}*' = \d_{\hat{I},\bar{J}-n}\cP_{\bar{J}}.
\ee
Therefore,
\be\label{Yprop}
Y_{\bar{I}\bar{J}} = Y_{\hat{I}\hat{J}} = 0,\quad
Y_{\hat{I}\bar{J}}
=Y_{\bar{J}\hat{I}}
=\d_{\hat{I},\bar{J}-n}.
\ee

Let us replace $dP$ in eq.\eq{SM5-mod} by $X$
and introduce Lagrange multiplier $\Pt$
which puts $dX = 0.$
The action becomes
\be\label{S-nlX}
\begin{split}
	S &= \hlf\int\bigg(\hlf X\w*' X-2 Q\w X - d\Pt\w X\\
		&\quad\qquad+\frac{d^6 x}{12}\sqrt{-g}U+Q\w R + F\w C_3 + 2C_6\bigg).
\end{split}
\ee
Euler-Lagrange equation for $X$ is given by
\be\label{X-EL}
X +2Q+*'d\Pt = 0.
\ee
As suggested by \cite{Lambert:2023qgs},
if we substitute $X$ from eq.\eq{X-EL} into eq.\eq{S-nlX},
we would obtain the action as in eq.\eq{SM5-mod}
with $P$ replaced by $\Pt.$
This is not the resulting action we are interested in.
So let us instead only substitute the components $\cP_{\bar{J}}X$
from eq.\eq{X-EL}:
\be
\cP_{\bar{J}}X =-2\cP_{\bar{J}}Q-\cP_{\bar{J}}*'d\Pt.
\ee
into eq.\eq{S-nlX}.
This gives
\be\label{S-elim-XbarJ}
\begin{split}
	S&=\hlf\int\bigg(2\cP_{\hat{I}}Q\w*'\cP_{\hat{I}}Q
	+2\cP_{\hat{I}}d\Pt\w*'\cP_{\hat{I}}Q\\
	&\qquad\qquad+\hlf\cP_{\hat{I}}d\Pt\w*'\cP_{\hat{I}}d\Pt
	+\hlf \cP_{\hat{I}}X\w*'\cP_{\hat{I}}X\\
	&\qquad\qquad-2*' \cP_{\hat{I}} Q\w \cP_{\hat{I}}X
	- Y_{\bar{J}\hat{I}}\cP_{\bar{J}} d\Pt\w \cP_{\hat{I}} X\\
	&\qquad\qquad+\frac{d^6 x}{12}\sqrt{-g}U+Q\w R+ F\w C_3 + 2C_6\bigg).
\end{split}
\ee

Note that if $\o$ is a differential $p$-form on $\cM_6$ and is independent on coordinates $x^\a$ of $\cC_{6-D}$, we have $d\o = \hat{d}\o \equiv dx^a\w\pa_a\o .$
In particular, if $\hat{d}\o = 0,$ then by expressing in index-notation and using the fact that
the $D$-dimensional spacetime is topologically trivial, it can be shown, for example by expressing in index notation, that there exists a $(p-1)$-form $\f$ on $\cM_6$ such that it is independent on $x^\a$
and that $\o = \hat{d}\f.$

The Euler-Lagrange equations for $\Pt$ are
\be
\hat{d}(2\cP_{\bar{J}}\cP_{\bar{J}} Q + *'\cP_{\hat{I}}\cP_{\hat{I}}d\Pt - \cP_{\hat{I}}\cP_{\hat{I}}X)
=0,
\ee
which imply
\be\label{Pt-EL-2}
2\cP_{\bar{J}}\cP_{\bar{J}} Q + *'\cP_{\hat{I}}\cP_{\hat{I}}d\Pt - \cP_{\hat{I}}\cP_{\hat{I}}X
=-\hat{d}P,
\ee
where we denote the arbitrary
$x^\a$-independent $2-$form on the $6$-dimensional spacetime
as $P.$ Let us apply $\cP_{\hat{I}}$
on eq.\eq{Pt-EL-2}. This gives
\be\label{XhatI-elim}
\begin{split}
	\cP_{\hat{I}}X
	&= \cP_{\hat{I}}\dh P.
\end{split}
\ee
By substituting eq.\eq{XhatI-elim} into eq.\eq{S-elim-XbarJ},
we obtain
\be\label{S-elim-X}
\begin{split}
	S&=\hlf\int\bigg(2\cP_{\hat{I}}Q\w*'\cP_{\hat{I}}Q
	-2*' \cP_{\hat{I}} Q\w \cP_{\hat{I}}(d P+d\Pt)\\
	&\qquad\qquad+\hlf\cP_{\hat{I}}d\Pt\w*'\cP_{\hat{I}}d\Pt
	+\hlf \cP_{\hat{I}}d P\w*'\cP_{\hat{I}}d P\\
	&\qquad\qquad
	+\frac{d^6 x}{12}\sqrt{-g}U+Q\w R+ F\w C_3 + 2C_6\bigg),
\end{split}
\ee
where we discard the integral of total derivative.

The next task is to eliminate $Q$ by solving its Euler-Lagrange equations and substituting back into the action \eq{S-elim-X}.
Let us argue that this is very difficult, if at all possible, to approach this directly.
In principle, the Euler-Lagrange equations of $Q$ would not be sufficient to completely determine $Q.$ One should also realise self-duality and non-linear self-duality conditions. In particular, one should solve eq.\eq{nonlin-M5} to determine $R$ in terms of $Q, g, C_3.$
The resulting expression of $Q$ is expected to be in terms of $P, \Pt, g, C_3.$ It is likely that this expression would be complicated so that it is not even clear if it can be expressible in a closed form.

With the insight gained from the analysis in section \ref{sec:ddim-circle}, one may approach this strategically. Let us first consider the Euler-Lagrange equations for $Q$ which can be expressed as \be\label{RhatI-QhatI}
\cP_{\hat{I}}R
=\cP_{\hat{I}}Q
+\hlf\cP_{\hat{I}}(dP+d\Pt).
\ee
Note that one can see from eq.\eq{RhatI-QhatI} that some projections of $F = Q-R+C_3$ turn out to be expressible as simple expressions of $P$ and $\Pt.$ That is
\be\label{HhatI}
\cP_{\hat{I}} F = -\hlf\cP_{\hat{I}}(dP + d\Pt) + \cP_{\hat{I}}C_3,
\ee
which when using the property eq.\eq{cPIdA}, the first term on the RHS of eq.\eq{HhatI} is an exact differential form. One may use eq.\eq{RhatI-QhatI} to express $R$ in terms of $Q$ then use
eq.\eq{nonlin-M5} to express $Q$ in terms of $\cP_{\hat{I}}F.$ The idea is that the collection 
$\{\cP_{\hat{I}}F\ |\ \hat{I} = 1,2,\cdots,n\}$ contains ten components of $F.$ The other ten components which are contained in the collection $\{\cP_{\bar{J}}F\ |\ \bar{J} = n+1,n+2,\cdots,2n\}$ can in principle be expressed in terms of $\cP_{\hat{I}}F$ by using non-linear self-duality relation of $F$. This means that eq.\eq{nonlin-M5} can be used to express
\be\label{HbarJ}
\cP_{\bar{J}}F
=\hat{V}_{\bar{J}}(\cP_{\hat{I}}F,g),
\ee
where $\hat{V}_{\bar{J}}(\cP_{\hat{I}}F,g)$ is a differential $3$-form which is a function of $\cP_{\hat{I}}F$ and $g.$ Then by using $F = Q-R+C_3$ along with eq.\eq{RhatI-QhatI}-\eq{HbarJ}, we obtain
\be\label{QhatI-aftersubs}
\cP_{\hat{I}}Q
=\hlf\cP_{\hat{I}} F + \hlf Y_{\hat{I}\bar{J}}*'\hat{V}_{\bar{J}}(\cP_{\hat{I}}F,g)
-\hlf\cP_{\hat{I}}(C_3+*'C_3).
\ee
Let us then substitute eq.\eq{RhatI-QhatI} and eq.\eq{QhatI-aftersubs} into the action \eq{S-elim-X}.
Note also that we may substitute the components $\cP_{\bar{J}}F$ from eq.\eq{HbarJ} into the expression of $U(F,g)$ leaving only the dependency on $\cP_{\hat{I}}F$ and $g.$ More explicitly, we have after substitution,
\be\label{U-aftersubs}
U(F,g)=\hat{U}(\cP_{\hat{I}}F,g).
\ee
The action \eq{S-elim-X} then becomes
\be\label{S-elim-R-2}
\begin{split}
	S&=\hlf\int\bigg(
	\ove{4}\cP_{\hat{I}}(dP-d\Pt)\w*'\cP_{\hat{I}}(dP-d\Pt)\\
	&\qquad\qquad+Y_{\hat{I}\bar{J}}\hat{V}_{\bar{J}}(\cP_{\hat{I}}F,g)\w \cP_{\hat{I}}F\\
	&\qquad\qquad
	+\frac{d^6 x}{12}\sqrt{-g}\hat{U}(\cP_{\hat{I}}F,g)\\
	&\qquad\qquad+\cP_{\hat{I}}\cP_{\hat{I}}(2F-C_3)\w C_3+2C_6 \bigg).
\end{split}
\ee
Let us make field redefinitions
\be
A = -\hlf (P+\Pt),\quad
A^{(s)} = \hlf(P-\Pt).
\ee
We may write 
\be
\cP_{\hat{I}}dA
=\dh\hat{A}_{\hat{I}},\quad
\cP_{\hat{I}}dA^{(s)}
=\dh\hat{A}^{(s)}_{\hat{I}}.
\ee
Therefore,
\be\label{S-reduced}
\begin{split}
	S&=\hlf\int\bigg(
	\dh\hat{A}^{(s)}_{\hat{I}}\w*'\dh\hat{A}^{(s)}_{\hat{I}}
	+Y_{\hat{I}\bar{J}}\hat{V}_{\bar{J}}(\Fh_{\hat{K}},g)\w \Fh_{\hat{I}}\\
	&\qquad\qquad+\frac{d^6 x}{12}\sqrt{-g}\hat{U}(\Fh_{\hat{I}},g)\\
	&\qquad\qquad
	+\cP_{\hat{I}}(2F_{\hat{I}}-\cP_{\hat{I}}C_3)\w C_3+2C_6\bigg),
\end{split}
\ee
where $\Fh_{\hat{I}}\equiv \dh\hat{A}_{\hat{I}}+\cP_{\hat{I}}C_3.$
After integrating out internal coordinates $x^\a$
the action \eq{S-reduced} physical fields  $\hat{A}_{\hat{I}}$ decoupled with free unphysical fields $\hat{A}^{(s)}_{\hat{I}}$. 
Generically, each of $\hat{A}_{\hat{I}}$
are collections of physical $0$-form, $1$-form, and $2$-form fields in the $D$ dimensional spacetime.

\subsection{Duality}
In subsection \ref{subsec:dual-D4-dD4}, we have seen that
there is a simple way to realise the duality between D4-brane and dual D4-brane. From the point of view of the Sen M5-brane action, the duality is realised by simply switching the roles of components of $P.$ In fact, it is natural to extend this idea to the cases of dimensional reduction to even lower dimensions.

Let us start by giving a quick summary to the result of subsection \ref{subsec:dimred} as follows.
Consider a dimensional reduction of the M5-brane action \eq{SM5-mod} to a $D$ dimensional spacetime. The appropriate choice of $2n$ projection operators satisfy eq.\eq{cPIconst}-\eq{Yprop} are also chosen. It turns out that only the first $n$ projection operators $\cP_1,\cP_2,\cdots,\cP_n$ explicitly appear in the reduced action \eq{S-reduced}, which let us denote as $S_{1,2,3,\cdots,n}.$

If we swap for example $\cP_1$ with $\cP_{n+1},$
the projection operators would still satisfy the properties
\eq{cPIconst}-\eq{Yprop}. Therefore, we may obtain the reduced action $S_{n+1,2,3,\cdots,n}$
which is explicitly described by the projection operators $\cP_{n+1}, \cP_2, \cP_3,\cdots,\cP_{n}.$

In fact, the theories $S_{1,2,3,\cdots,n}$ and $S_{n+1,2,3,\cdots,n}$
are related by duality transformation.
We first note that the action $S_{1,2,3,\cdots,n}$ is equivalent to the action \eq{S-elim-X}. Focusing on terms involving $\cP_{1}Q, \cP_{1}dP, \cP_{1}d\Pt,$ we may express the action \eq{S-elim-X} as
\be\label{S-elim-X-hat1}
\begin{split}
	&S_{1,2,3,\cdots,n}\\
	&=\hlf\int\bigg(2\cP_{1}Q\w*'\cP_{1}Q\\
	&\qquad\qquad-2*' \cP_{1} Q\w \cP_{1}(d P+d\Pt)\\
	&\qquad\qquad+\hlf\cP_{1}d\Pt\w*'\cP_{1}d\Pt\\
	&\qquad\qquad+\hlf \cP_{1}d P\w*'\cP_{1}d P\bigg) +S',
\end{split}
\ee
where
\be
\begin{split}
	S'&=\hlf\int\bigg(\sum_{\hat{I}=2}^n2\cP_{\hat{I}}Q\w*'\cP_{\hat{I}}Q\\
	&\quad\qquad-\sum_{\hat{I}=2}^n2*' \cP_{\hat{I}} Q\w \cP_{\hat{I}}(d P+d\Pt)\\
	&\quad\qquad+\sum_{\hat{I}=2}^n\hlf\cP_{\hat{I}}d\Pt\w*'\cP_{\hat{I}}d\Pt\\
	&\quad\qquad+\sum_{\hat{I}=2}^n\hlf \cP_{\hat{I}}d P\w*'\cP_{\hat{I}}d P\\
	&\quad\qquad
	+\frac{d^6 x}{12}\sqrt{-g}U+Q\w R+ F\w C_3 + 2C_6\bigg).
\end{split}
\ee

Note that by using the property
\eq{cPIdA} and the fact that $P$ and $\Pt$ are independent from $x^\a,$ we may express
\be
\cP_1 dP = \dh P_1,\qquad
\cP_1 d\Pt = \dh \Pt_1.
\ee
In the action \eq{S-elim-X-hat1}, let us set $\F\equiv \dh P_1, \Phit\equiv \dh \Pt_1$ and introduce Lagrange multiplier to set $\dh \F = \dh\Phit = 0$ and $\cP_1 \F = \F, \cP_1\Phit = \Phit.$ So
\be\label{S-elim-X-hat1-equiv}
\begin{split}
	&S_{1,2,3,\cdots,n}\\
	&=\hlf\int\bigg(2\cP_{1}Q\w*'\cP_{1}Q
	-2*' \cP_{1} Q\w (\F+\Phit)\\
	&\qquad\qquad+\hlf\Phit\w*'\Phit
	+\hlf \F\w*'\F\\
	&\qquad\qquad-\cP_{n+1}dK\w\F-\cP_{n+1}d\Kt\w\Phit\bigg) + S',
\end{split}
\ee
where $K$ and $\Kt$ are independent from $x^\a.$
Euler-Lagrange equations for $\F$ and $\Phit$ are
\be
\F = -2\cP_1Q - *'\cP_{n+1}dK,\quad
\Phit = -2\cP_1Q - *'\cP_{n+1}d\Kt.
\ee
Substituting into the action \eq{S-elim-X-hat1-equiv} gives
\be\label{S-elim-X-hat1-equiv-dual}
\begin{split}
	&S_{1,2,3,\cdots,n}\\
	&=\hlf\int\bigg(2\cP_{n+1}Q\w*'\cP_{n+1}Q\\
	&\qquad\qquad-2*' \cP_{n+1} Q\w \cP_{n+1}(d K+d\Kt)\\
	&\qquad\qquad+\hlf\cP_{n+1}d\Kt\w*'\cP_{n+1}d\Kt\\
	&\qquad\qquad+\hlf \cP_{n+1}d K\w*'\cP_{n+1}d K\bigg) + S',
\end{split}
\ee
which after field redefinitions, it can easily be seen that
\be
S_{1,2,3,\cdots,n}
=S_{n+1,2,3,\cdots,n},
\ee
as required.

The analysis above indeed generalises the analysis of
section \ref{sec:ddim-circle},
in which $D=5.$ So if we define $\cP_1, \cP_2$ such that
\be\label{cP-D4}
\begin{split}
	&\cP_1(dx^\m\w dx^\n\w dx^\r) = 
	3\d^{\m\n\r}_{ij5}dx^i\w dx^j\w dx^5,\\
	&\cP_2(dx^\m\w dx^\n\w dx^\r) = 
	\d^{\m\n\r}_{ijk}dx^i\w dx^j\w dx^k,
\end{split}
\ee
then the physical part of the action \eq{S-reduced} describes a D4-brane. But if we define 
\be\label{cP-dD4}
\begin{split}
	&\cP_1(dx^\m\w dx^\n\w dx^\r) = 
	\d^{\m\n\r}_{ijk}dx^i\w dx^j\w dx^k,\\
	&\cP_2(dx^\m\w dx^\n\w dx^\r) = 
	3\d^{\m\n\r}_{ij5}dx^i\w dx^j\w dx^5,
\end{split}
\ee
then the physical part of the action \eq{S-reduced} describes a dual D4-brane.

By similar considerations, any pair $(\cP_k, \cP_{k+n}), k=1,2,\cdots,n$ of projection operators can be swapped and that it is also possible to swap more than one pair at a time. The resulting action is dual to the original action.
For simplicity, let us discuss an example case where the action is quadratic.
In this case, the action \eq{S-reduced} is
\be\label{S-reduced-quad}
\begin{split}
S
=&\hlf\int\bigg(\dh\hat{A}^{(s)}_{\hat{I}}\w*'\dh\hat{A}^{(s)}_{\hat{I}}
	+Y_{\hat{I}\bar{J}}\hat{V}_{\bar{J}}(\Fh_{\hat{K}},g)\w \Fh_{\hat{I}}\\
	&\qquad+\cP_{\hat{I}}(2\Fh_{\hat{I}}-\cP_{\hat{I}}C_3)\w C_3+2C_6\bigg),
\end{split}
\ee
where the form of $\hat{V}_{\bar{J}}(\Fh_{\hat{I}},g)$
will be given on a case by case basis by rearranging the 6d $*-$self-duality condition of $F$.

In particular, let us consider an example case in which $D=4,$
the compact space is a torus, and the 6d metric is $g = g_4\oplus g_2.$
Let us pick the projection operators as follows
\be\label{Deq4-proj}
\begin{split}
	\cP_1(dx^\m\w dx^\n\w dx^\r)
	&=6\d^{\m\n\r}_{i45}dx^i\w dx^4\w dx^5,\\
	\cP_2(dx^\m\w dx^\n\w dx^\r)
	&=3\d^{\m\n\r}_{ij4}dx^i\w dx^j\w dx^4,\\
	\cP_3(dx^\m\w dx^\n\w dx^\r)
	&=\d^{\m\n\r}_{ijk}dx^i\w dx^j\w dx^k,\\
	\cP_4(dx^\m\w dx^\n\w dx^\r)
	&=3\d^{\m\n\r}_{ij5}dx^i\w dx^j\w dx^5.
\end{split}
\ee
Furthermore, let us consider a consistent truncation by setting
\be
\Fh_1 = \Fh_3 = \cP_1 C_3 = \cP_3 C_3 = 0,
\ee
or in component form
\be
F_{i45} = F_{ijk} = C_{i45} = C_{ijk} = 0.
\ee
This example case is exactly what studied before in \cite{Sen:2019qit}
in which $g_4 = \h_4$ is the flat 4d metric. Here, however, we shall generalise this by taking $g_4$ to be a general 4d metric which depends on the 4d coordinates $x^i.$
The $*$-self-duality condition of $F$ in 6d gives
\be
F^{ij4}
=\ove{g_{44}}\lrbrk{g^{mi}g^{nj} + g_{45}\ove{2\sqrt{-g}}\e^{ijmn}}F_{mn4},
\ee
where $F_{mn4} = 2\pa_{[m} A_{n]4} + C_{mn4}.$ Therefore, the physical part of the action \eq{S-reduced-quad} is
\be\label{S-D3-quad}
\begin{split}
	&S_{phys}\\
	&=\int d^4 x\bigg(-\ove{4}\frac{\sqrt{g_{44}g_{55}-g_{45}^2}}{g_{44}}\sqrt{-g_4}F_{ij4}F_{mn4}g^{im}g^{jn}\\
	&\quad\qquad-\ove{8}\frac{g_{45}}{g_{44}}\e^{mnij}F_{ij4}F_{mn4}\bigg)\\
	&+
	\int d^4 x\bigg(-\ove{4} F_{ij4}C_{mn5}\e^{ijmn}
		-\ove{8}C_{ij4}C_{mn5}\e^{ijmn}\\
	&\quad\qquad-\ove{4!}C_{ijmn45}\e^{ijmn}\bigg),
\end{split}
\ee
where $\e^{ijmn}$ is a 4d Levi-Civita tensor which is defined as
$\e^{ijmn} = \e^{ijmn45},$
and for simplicity, we set $\int dx^4 = \int dx^5 = 1.$
By making the swap $\cP_2\lrar\cP_4,$ we obtain the dual action
\be\label{S-D3-quad-dual}
\begin{split}
	&\St_{phys}\\
	&=\int d^4 x\bigg(-\ove{4}\frac{\sqrt{g_{44}g_{55}-g_{45}^2}}{g_{55}}\sqrt{-g_4}F_{ij5}F_{mn5}g^{im}g^{jn}\\
	&\qquad\qquad+\ove{8}\frac{g_{45}}{g_{55}}\e^{mnij}F_{ij5}F_{mn5}\bigg)\\
	&\quad+
	\int d^4 x\bigg(\ove{4} F_{ij5}C_{mn4}\e^{ijmn}+\ove{8}C_{ij4}C_{mn5}\e^{ijmn}\\
	&\qquad\qquad-\ove{4!}C_{ijmn45}\e^{ijmn}\bigg).
\end{split}
\ee
The action \eq{S-D3-quad} and its dual \eq{S-D3-quad-dual}
transform to each other under the S-duality transformation
\be\label{g2}
g_{44}\mapsto g_{55},\quad
g_{45}\mapsto -g_{45},\quad
g_{55}\mapsto g_{44},
\ee
\be\label{B2C2}
\begin{split}
	&F_{ij4}\mapsto - F_{ij5},\quad
	F_{ij5}\mapsto F_{ij4},\\
	&C_{ij4}\mapsto -C_{ij5},\quad
	C_{ij5}\mapsto C_{ij4}.
\end{split}
\ee
Note that the transformation \eq{g2} is equivalent to
\be\label{modular}
\t\mapsto -\ove{\t},
\ee
where
\be
\t = \frac{-g_{45}+i\sqrt{g_{44}g_{55}-g_{45}^2}}{g_{44}}.
\ee
The transformation \eq{modular} is
what expected from the conformal symmetry of the
6d action of self-dual 3-form
\cite{Witten:2009at}.
Furthermore, in the context of D3-brane, $\t$ encodes axion and dilation as $\t = C_0 + ie^{-\f},$ whereas $C_{ij4}$ and $C_{ij5}$ are identified with $B_2$ and $C_2.$ The transformations \eq{g2}-\eq{B2C2} is precisely the S-duality transformation for D3-brane.

It is clear that the reduced theories \eq{S-D3-quad}-\eq{S-D3-quad-dual} we have obtained is expressed entirely in terms of standard fields. This is possible largely due to that we have chosen appropriate projection operators, in particular the ones which project to $ij4$ or $ij5$ components. On the other hand, for the choice studied in \cite{Sen:2019qit}, in which the components $\ih\jh\a$ where $\ih,\jh\in\{1,2,3\}, \a\in\{4,5\}$ are chosen, it is not yet clear whether the Lagrangian can be expressed entirely in terms of standard fields.

Note that due to the expected form of the reduced theory under dimensional reduction and how it transforms under S-duality,
it could be expected that if one considers double dimensional reduction of the complete Sen M5-brane action \eq{SM5-mod} on a torus, then one could obtain the complete D3-brane action on a circle. Indeed, explicit checks are required. If this succeeds, one may then compare and contrast the approach with its counterpart \cite{Berman:1998va} given in the PST formalism.

It is remarkable that the duality can easily be obtained simply by appropriately swapping the projection operators. In fact, since the analysis only involves mainly the field $P$ and projection operators, similar analysis for the duality of dimensional reduction of self-dual $(2p+1)$-forms in $4p+2$ dimensions in Sen formalism can easily be performed.
This way that dualities of dimensionally reduced theories are realised in the Sen formalism is another property which makes the Sen formalism appealing. Of course,
further efforts are still required to eliminate $Q$ from each action to arrive at the final form.
\section{Conclusion and Discussion}\label{sec:conclusion}
In this work, we have studied double dimensional reduction of the Sen M5-brane action \eq{SM5-mod}. In particular, we have explicitly shown that double dimensional reduction of the Sen M5-brane action on a circle indeed gives, depending on how the projection operators are chosen, the complete D4-brane action or the complete dual D4-brane action. The procedure we have followed is essentially just a slight modification of that given in \cite{Andriolo:2020ykk}. It is remarkable that although the Sen M5-brane action couples to gravity in a complicated way and that there are fields which do not transform in the standard way under diffeomorphism, it seems that dimensional reduction can be carried out relatively simply giving rise to reduced actions which describe dynamics of fields which transform in a standard way under diffeormorphism.

We have also outlined an extension of dimensional reduction on other spaces. The analysis suggests that provided that one picks projection operators which satisfy eq.\eq{cPIconst}-\eq{Yprop}, the dimensional reduction of the Sen M5-brane action gives rise to the reduced action \eq{S-reduced} which contains standard fields.

Note that one aspect which makes the procedure of dimensional reduction simple despite the aforementioned complications of the Sen M5-brane action is that it allows one to easily integrating out $Q.$ In particular,
it avoids the need to explicitly express $R$ in terms of $Q, g, C_3.$ As a future work, it would be interesting to see whether it is possible to extend and modify the procedure to show explicitly that the Sen M5-brane action is equivalent to the M5-brane actions in PST formalism. We anticipate that if this is at all possible, the procedure should also allow one to integrate out $Q$ in a simple way.

We have found another remarkable feature of Sen M5-brane action that dualities of dimensional reduced action seem to be encoded within the Sen M5-brane action. For example, as demonstrated explicitly in this paper, in the viewpoint of string theory, it would take some efforts to show that the D4-brane and dual D4-brane actions are dual to each other. On the other hand, from the viewpoint of the Sen M5-brane action, the D4-brane and dual D4-brane actions can be put in a similar form. This can be seen by comparing eq.\eq{S-pred-D4-3} and eq.\eq{S-pred-dualD4-2}, which can be transformed to each other simply by swapping the projection operators between eq.\eq{cP-D4} and eq.\eq{cP-dD4}. We have also shown that this swapping is a result of dualisation.

We have outlined a generalisation to dimensional reduction on other manifolds. As long as the projection operators can be given to satisfy \eq{cPIconst}-\eq{Yprop}, the process will give rise to the reduced action \eq{S-reduced} whose physical part describe fields which transform under standard diffeomorphism. Apart from the D4-brane and dual D4-brane actions, it can be expected that the action \eq{S-reduced} should also give rise to other complete actions in string theory and M-theory for example D3-brane, D2-brane, D1-brane, F1-brane, M2-brane actions
and whether S-dualities in these cases are the results of simple swapings of the projection operators. For completeness, explicit checks should be made in each case.

The analysis for dimensional reduction on other spaces as given in section \ref{sec:dimred-otherspaces} heavily relies on the assumption that the projection operators should satisfy eq.\eq{cPIconst}-\eq{Yprop}. It would be interesting to see whether the analysis can be extended to the cases where some of the properties \eq{cPIconst}-\eq{Yprop} are generalised.
In any case, it would also be interesting to consider dimensional reduction of the Sen M5-brane action on other spaces. In particular, one might consider the dimensional reduction of the Sen M5-brane action on Riemann surface and K3. This is to see non-linear extension to the result of \cite{Andriolo:2020ykk}.

It could also be expected that other kinds of dimensional reduction for the Sen M5-brane action could also easily be carried out. In particular, the null dimensional reduction on a (2,0) theory in the Sen formalism has been carried out in \cite{Lambert:2020scy}. So one might try to use a similar method to study the null reduction of the Sen M5-brane action.

Although the analysis in section \ref{sec:dimred-otherspaces} is given explicitly for the six-dimensional Sen M5-brane action, the extension especially for the quadratic version which describes self-dual $(2p+1)$-form on ${4p+2}$ dimensional spacetime can easily be done. In particular, for the case $p = 2,$ the Sen quadratic action describes type IIB supergravity. So it would also be interesting to study how the dimensional reduction of type IIB supergravity is realised in Sen formalism.
\section*{Acknowledgements}
We would like to thank Sheng-Lan Ko
for discussions. A.P. is supported by a Development and Promotion of Science and Technology Talents Project (DPST) scholarship from the Royal Thai Government. This version of the article has been accepted for publication, after peer review 
but is not the Version of Record and does not reflect post-acceptance improvements, or any 
corrections. The Version of Record is available online at: http://dx.doi.org/10.1140/epjc/s10052-023-11892-2

\vspace{1cm}

\noindent\textbf{Data Availability Statement} No data was analysed or generated in this work.

\end{document}